The longevity of water ice on Ganymedes and Europas around migrated giant planets

Owen R. Lehmer[1], David C. Catling[1], Kevin J. Zahnle[2]
[1] Dept. of Earth and Space Sciences / Astrobiology Program, University of Washington, Seattle, WA
[2] NASA Ames Research Center, Moffett Field, CA

**ABSTRACT**

The gas giant planets in the Solar System have a retinue of icy moons, and we expect giant exoplanets to have similar satellite systems. If a Jupiter-like planet were to migrate toward its parent star the icy moons orbiting it would evaporate, creating atmospheres and possible habitable surface oceans. Here, we examine how long the surface ice and possible oceans would last before being hydrodynamically lost to space. The hydrodynamic loss rate from the moons is determined, in large part, by the stellar flux available for absorption, which increases as the giant planet and icy moons migrate closer to the star. At some planet-star distance the stellar flux incident on the icy moons becomes so great that they enter a runaway greenhouse state. This runaway greenhouse state rapidly transfers all available surface water to the atmosphere as vapor, where it is easily lost from the small moons. However, for icy moons of Ganymede's size around a Sun-like star we found that surface water (either ice or liquid) can persist indefinitely outside the runaway greenhouse orbital distance. In contrast, the surface water on smaller moons of Europa's size will only persist on timescales greater than 1 Gyr at distances ranging 1.49 to 0.74 AU around a Sun-like star for Bond albedos of 0.2 and 0.8, where the lower albedo becomes relevant if ice melts. Consequently, small moons can lose their icy shells, which would create a torus of H atoms around their host planet that might be detectable in future observations.

**1. INTRODUCTION**

One of the major results of exoplanet discoveries is that giant planets migrate (Chambers 2009). This was first deduced from hot Jupiters, and although these are found around 0.5-1% of Sun-like stars (Howard 2013), hot Jupiters are not the only planets to migrate and giant planet migration is likely widespread. Indeed, such migration probably occurred in the early solar system (Tsiganis et al. 2005).

All the giant planets in the solar system have a collection of icy moons. We expect that similar exomoons orbit giant exoplanets and that these moons would likely migrate along with their host planet. If a giant exoplanet were to migrate toward its parent star, icy moons could vaporize, similar to comets approaching the Sun, and develop atmospheres. If such a giant planet and icy moons were to form in the habitable zone of a star (or migrate there shortly after formation) the high XUV flux from the young star would rapidly erode the atmospheres of moons many times the mass of Ganymede (). In addition, they could melt and maintain liquid surfaces as they migrate inwards, which could be potentially habitable environments. Such a moon would have an atmosphere primarily controlled by the vapor equilibrium set by the surface temperature and the rate of hydrodynamic escape to space. As such, the longevity of the water shell and atmosphere will depend primarily on the distance to the host star and the exomoon radius and mass.

Several such bodies exist in the solar system, where the atmospheric thickness is determined by vapor equilibrium with a condensed phase, i.e. the Clausius-Clapeyron relation for the relevant volatile. Let us call such atmospheres *Clausius-Clapeyron (C-C) atmospheres*. The $N_2$ atmospheres on both Triton and Pluto are examples of C-C atmospheres, where the surface vapor pressure is in equilibrium with the $N_2$ surface ice at the prevailing temperature for each body. The present Martian atmosphere is another C-C atmosphere since the polar $CO_2$ ice caps at ~148 K buffer the atmosphere to ~600 Pa surface pressure (Leighton & Murray 1966) (see (Kahn 1985) for an explanation over geologic timescales).

For an icy exomoon migrating toward its parent star, the atmospheric water vapor will be controlled by the availability of surface water and temperature. Very deep ice and ice-covered oceans are possible on these moons given that water can account for ~5-40% of the bulk mass of icy moons in the solar system (Schubert et al. 2004). However, the small mass of exomoons and relatively high stellar flux as the exomoon migrates toward the star makes water vapor susceptible to escape. Assuming exomoons are of comparable size to the moons found in the solar system, this study looks at the end-member case of how rapidly a pure water vapor atmosphere will be lost hydrodynamically during exomoon migration. The migration of exomoons is essential if icy moons of Ganymede's size are to retain their surface water for more than 1 Gyr in the habitable zone of a Sun-like star. If a Ganymede-like icy moon formed in the habitable zone of a star (or migrated there shortly after formation) the high XUV flux from the young star could rapidly erode its atmosphere (Heller, Marleau, & Pudritz 2015; Lammer et al. 2014). Therefore, this study looks at the longevity of surface water on icy moons that migrate toward their host star after this period of intense XUV-driven hydrodynamic escape.

Hydrodynamic escape is a form of pressure-driven thermal escape where the upper levels of an atmosphere become heated and expand rapidly, accelerate through the speed of sound, and escape to space *en masse* (Hunten 1990). An important process in atmospheric evolution, hydrodynamic escape likely occurred during the formation of the terrestrial atmospheres (Kramers & Tolstikhin 2006; Kuramoto, Umemoto, & Ishiwatari 2013; Pepin 1997; Tolstikhin & O'Nions 1994). Moreover, hydrodynamic escape has been observed on exoplanets such as the gas giant HD 209458b, which orbits a Sun-like star at 0.05 AU and has hot H atoms beyond its Roche lobe, presumably deposited there by hydrodynamic escape (Linsky et al. 2010; Vidal-Madjar et al. 2004). The closer a body is to its parent star, the more effective the hydrodynamic escape, and the smaller the body, the more easily an atmosphere is lost (e.g., Zahnle & Catling (2017)).

The longevity of an atmosphere and icy shell will depend primarily on temperature, set in large part by the stellar flux available for absorption. As an ice covered exomoon moves towards its parent star, heating will cause more water vapor to enter the atmosphere, hastening the loss rate. In addition, this water vapor will provide a greenhouse effect, further warming the moon. At a certain exomoon-star distance the water vapor atmosphere will impose a runaway greenhouse limit on the outgoing thermal infrared (IR) flux from the exomoon. If the absorbed stellar flux exceeds this limit, the exomoon surface will heat rapidly until all available water is in the atmosphere as vapor. This limit represents the distance at which all surface water will be transferred to the atmosphere where it will be rapidly lost.

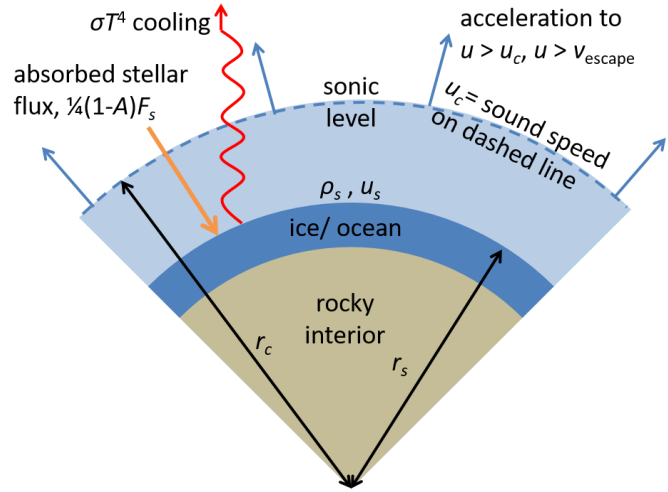

**Figure 1.** Conceptual visualization of the 1D hydrodynamic escape model. The incoming absorbed stellar flux, given by $\frac{1}{4}(1-A)F_s$, heats the exomoon surface, for Bond albedo $A$, and stellar flux $F_s$. The exomoon will remain in thermal equilibrium by evaporating water vapor, losing mass via hydrodynamic escape, and radiating in the thermal infrared. The thermal infrared radiation is given by $\sigma T^4$ ($\sigma$ is the Stefan-Boltzmann constant, $T$ is the surface temperature) in a blackbody approximation. The outward radial flow velocity, $u$, increases monotonically until it surpasses the isothermal speed of sound, $u_c$, at the critical radius, $r_c$, where the gas is still collisional. Beyond the sonic level, $u$ continues to rise and soon surpasses the escape velocity $v_{escape}$. The exomoon surface is at radius $r_s$ with atmospheric near-surface density $\rho_s$ and outward radial surface velocity $u_s$. See equation (2) for the global energy balance.

## 2. METHODS

We consider three cases of hydrodynamic escape: (**2.1**) an isothermal atmosphere where the atmospheric temperature is set by incoming stellar flux and equal to the effective temperature; (**2.2**) a vapor saturated atmosphere where the temperature and humidity profiles of the entire atmosphere are dictated by the C-C relation; and (**2.3**) an isothermal atmosphere similar to (**2.1**) but the surface temperature, and the isothermal atmospheric temperature, are increased from the effective temperature by the total greenhouse warming of the water vapor atmosphere. We chose the isothermal and C-C cases because they present upper and lower limits on the rates of hydrodynamic escape, respectively, as described below. Figure 1 provides a conceptual picture of the model.

For a pure water vapor atmosphere around a Sun-like star, an isothermal atmosphere at the effective temperature represents the greatest possible temperature at the top of atmosphere to drive hydrodynamic escape. Water vapor radiates in the IR more efficiently than it absorbs sunlight, so the radiative-convective temperature for a pure water vapor atmosphere will be less (Pierrehumbert 2010; Robinson & Catling 2012). As such, the isothermal atmospheric approximation provides an upper bound on the atmospheric loss rate. In contrast, the lowest possible temperature to drive hydrodynamic escape is the saturated case, where the temperature and pressure at all heights are set by the C-C relation, which is defined by

$$T(P) = \frac{T_0}{1 - \ln(P/P_0)(RT_0/L_v)} \quad (1)$$

for reference temperature $T_0$ at reference pressure $P_0$, where $R$ is the universal molar gas constant, and $L_v$ is the latent heat of vaporization for water (e.g., Pierrehumbert (2010), p.100). The surface temperature is assumed to be in equilibrium with the incoming stellar flux, cooling associated with mass loss via hydrodynamic escape, and latent heat of evaporation. If the temperature were to decrease with altitude faster than the C-C relationship, the water vapor would condense out resulting in a C-C curve that, when extrapolated, would result in a surface temperature no longer in equilibrium with incoming stellar flux and escape. Therefore, the hydrodynamic loss rate of a pure water vapor atmosphere is bounded by the isothermal and saturated cases, which we will now consider in turn.

**2.1. Isothermal Case**

In the blackbody approximation, radiative cooling is given by $\sigma T^4$ at isothermal temperature $T$, allowing a straightforward formulation of escape versus radiative cooling. As such, the first order global energy balance for an icy exomoon is between incoming stellar flux versus the energy flux lost to vaporizing the water, lifting molecules out of the gravity well, and radiative cooling, i.e.,

$$\underbrace{\frac{1}{4}(1-A)F_s}_{\text{absorbed stellar flux}} = \underbrace{\left(\frac{GM}{r_s} + L_v\right)\rho_s u_s}_{\text{mass loss flux}} + \underbrace{\sigma T^4}_{\text{radiative cooling}} \qquad (2)$$

Here $F_s$ is the incoming solar flux available for absorption, $A$ is the Bond albedo, $\sigma$ is the Stefan-Boltzmann constant, $r_s$ is the surface radius where the atmospheric density is $\rho_s$, such that $\rho_s u_s$ is the mass flux given an outward radial flow velocity at the surface $u_s$, $G$ is the gravitational constant, and $M$ is the mass of the exomoon.

In addition to the C-C relationship [equation (1)], three equations are needed to derive the steady state, hydrodynamic atmospheric loss in the isothermal approximation (e.g., Catling & Kasting (2017), Ch. 5). The first is steady state mass continuity, given by

$$\frac{\partial}{\partial r}\left(r^2 \rho u\right) = 0 \qquad (3)$$

where $r$ is the radial distance from the planet's center, $u$ is the outward radial flow velocity, and $\rho$ is the atmospheric density. Steady state momentum conservation is expressed as

$$u\frac{\partial u}{\partial r} + \frac{1}{\rho}\frac{\partial p}{\partial r} = g \qquad (4)$$

with gravity $g = -GM/r^2$ and pressure $p$. Finally, the equation for energy balance is given by equation (2). Combining equations (1), (2), (3), and (4) an analytic expression for the isothermal atmospheric mass loss rate in kg s$^{-1}$ is given by (see Appendix A for the derivation):

$$\dot{M} = \pi \rho_s \frac{G^2}{u_0^3} M^2 \exp\left[\frac{3}{2} - \frac{G}{u_0^2}\left(\frac{3}{4\pi\rho_m}\right)^{-\frac{1}{3}} M^{\frac{2}{3}}\right] \quad (5)$$

where $u_0$ is the isothermal sound speed given by $u_0^2 = kT/m$ with Boltzmann constant $k$ and mean molecular weight $m$, and $\rho_m$ is the mean density of the exomoon (assumed 2 g cm$^{-3}$). The atmospheric surface density, $\rho_s$, is set the by C-C equation for the saturation vapor pressure of water at the prevailing temperature.

For time averaged mass loss rate, $\bar{\dot{M}}$, the lifetime of the exomoon surface water is given by

$$\tau_{Water} = \frac{M_{Water}}{\bar{\dot{M}}} \quad (6)$$

For an upper limit, we assume the total mass of water present on the exomoon surface, $M_{Water}$, is 40% of the bulk mass. However, even if 5% water were used [the lower limit for Europa (Schubert, et al. 2004)] from equation (6) we can see that it would translate to a change in $\tau_{Water}$ by a factor of 8, compared to 40% water. From equation (5) we see that $\dot{M}$, and hence $\tau_{Water}$, has an exponential dependence on mass, so we would anticipate that the difference between 5% and 40% water is not the major factor determining $\tau_{Water}$, which is borne out by our results. In addition, if substantial water vapor is lost the bulk density of the moon, $\rho_m$, may increase over time. However, from equation (5) we see that the exponential term scales like $-\rho_m^{1/3} M^{2/3}$ with $M^{2/3}$ largely determining the loss rate so the sensitivity to $\rho_m$ is small.

It is important to note that in equation (5) we have assumed the mass loss rate, $\dot{M}$, is sufficiently small that energy balance is dominated by radiative loss. This is indeed the case for exomoons of interest in this paper, where the low temperature water vapor atmospheres last for more than 1 Gyr. The surface pressures are well below ~500 Pa until the runaway greenhouse limit is reached. For bodies with rapid hydrodynamic escape the numerical approach defined in Appendix A is appropriate.

## 2.2. Saturated Temperature Profile Case

The saturated case is derived from the same equations as the isothermal case [equations (1), (2), (3), and (4)] but temperature is allowed to change with altitude. The temperature at the critical point at radius $r_c$ in Figure 1 (where the isothermal sound speed $u_0$ equals the radial escape speed) is set such that numerically integrating equations (1), (2), (3), and (4) from the critical point to the surface will result in a surface temperature equivalent to that in equilibrium with incoming solar flux taking into account the evaporative cooling (see Appendix A for details). Once the critical temperature is known the radial outflow velocity is readily calculated and thus the mass loss rate.

## 2.3. Isothermal Case with Greenhouse Effect Considered

In **Case (2.1)** we let the isothermal atmospheric temperature be set by just the incoming stellar flux and thus be equal to the effective temperature. However, for a thick water vapor atmosphere the surface will be heated by the greenhouse effect of the overlying atmosphere. In this case, we still used an isothermal atmosphere approximation but increased the atmospheric temperature by the total greenhouse warming of the atmosphere at the surface. The larger isothermal atmospheric temperature under this regime will increase the hydrodynamic loss rate compared to **Case (2.1)**.

To account for the atmospheric greenhouse effect, we used a gray, radiative, plane-parallel approximation where the total gray atmospheric optical depth in the thermal infrared at the surface is given by

$$\tau = \frac{\kappa_{ref} P^2}{2 g P_{ref}} \tag{7}$$

for mass absorption coefficient $\kappa_{ref}$ at pressure $P_{ref}$ and surface pressure $P$ where pressure broadening causes the $P^2$ dependency of the optical depth (Catling & Kasting 2017, p.381). Here we used $\kappa_{ref} = 0.05 \, \text{m}^2 \, \text{kg}^{-1}$ and $P_{ref} = 10^4$ Pa (from Catling & Kasting (2017), Ch. 13). Having $\tau \propto P^2$ in equation (7) is appropriate for thick atmospheres, which is the case when the runway greenhouse limit is approached. For thin atmospheres $\tau \propto P$ is appropriate (Catling & Kasting 2017, p.382). In this study, the surface pressures are in the low-pressure regime (less than ~500 Pa) until the runaway limit is reached. However, the difference between $\tau \propto P^2$ and $\tau \propto P$ in equation (7) is small at such low pressures where the total greenhouse warming is less than a few K until the runaway limit is reached. Setting $\tau \propto P$ for such low-pressure moons in equation (7) has negligible impact on the calculated mass loss rate so we approximate the optical depth of all atmospheres in this study with $\tau \propto P^2$. Once the total optical depth of the atmosphere is known from equation (7), the first order global energy balance is given by (see Appendix B for derivation)

$$\frac{1}{4}(1-A) F_s \left(1 + \frac{\tau}{2}\right) = \left(\frac{GM}{r_s} + L_v\right) \rho_s u_s + \sigma T_s^4 \tag{8}$$

and from equations (3) and (4) we derived an expression for $\rho_s u_s$ (see Appendix A)

$$\rho_s u_s = \rho_s u_0 \left(\frac{r_c}{r_s}\right)^2 \exp\left[-\frac{1}{2} + \frac{GM}{u_0^2}\left(\frac{1}{r_c} - \frac{1}{r_s}\right)\right] \tag{9}$$

Equations (1), (7), (8), and (9) were solved simultaneously to find $T$ and $u_s$, with $\rho_s$ being given by the ideal gas law. The mass loss rate is then calculated by

$$\dot{M} = 4\pi \rho_s u_s r_s^2 \tag{10}$$

Using equation (10) the time averaged loss rate is calculated and surface water lifetime is then obtained via equation (6).

It is possible that no physically meaningful solution exists to equations (1), (7), (8), and (9). When the initial surface temperature, and therefore surface pressure, is large (above ~260 K for this model), the optical depth given by equation (7) will be significant. This will cause an increase in surface temperature further increasing the surface pressure and thus the optical depth of the atmosphere. The positive feedback between temperature, pressure, and optical depth will cause equations (1), (7), (8), and (9) to have no valid solution if the initial surface temperature, set by the incoming stellar flux, is large. The exomoon-star distance where this positive feedback results in no solution is the runaway greenhouse limit, and it is akin the runaway limit found by Ingersoll (1969).

For all three model scenarios, we considered a pure water vapor atmosphere above a surface water reservoir. We looked at icy exomoons with masses ranging from 0.005 to 0.04 Earth masses between 0.9 and 2.0 AU from a Sun-like star. This mass range includes bodies slightly smaller than Europa (0.008$M_{Earth}$), and slightly larger than Ganymede (0.025$M_{Earth}$). We set the Bond albedo to 0.2 for each run. We chose a Bond albedo of 0.2 for two reasons, the first is that it approximately represents the lower bound for icy moon Bond albedos in the solar system (Buratti 1991; Howett, Spencer, & Pearl 2010). In addition, a Bond albedo of 0.2 approximates the albedo of open ocean with partial cloud cover (Goldblatt 2015; Leconte et al. 2013). Should an icy moon form surface oceans, the 0.2 Bond albedo gives us the best representation when calculating water longevity.

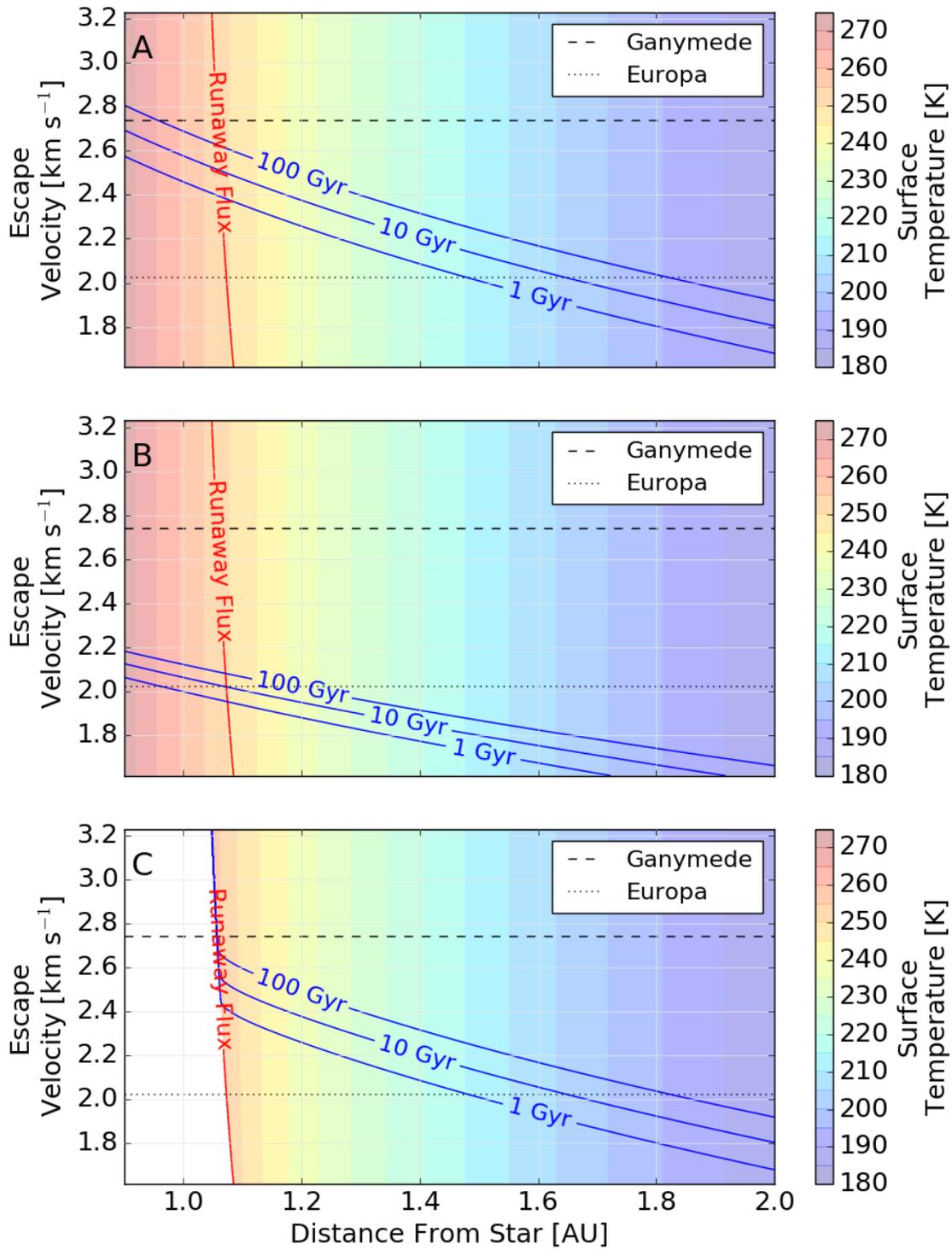

**Figure 2.** For all three plots, the red curve represents the *Runaway Flux* where the icy moon will be close enough to the star that a runaway greenhouse occurs. The blue curves represent contours of surface water lifetime (plotted in Gyr). The surface temperatures of the moons are shown by the colored background. In plots A and B the surface temperature corresponds to the effective temperature. In plot C the colored background shows the surface temperature, beyond the runaway limit distance, accounting for the effect of the water vapor greenhouse. Comparing the surface temperatures in plot C to those in A and B the water vapor greenhouse is negligible except very close to the runaway limit distance. The rate of hydrodynamic escape depends on both the mass and radius of the exomoon, as such we plot escape velocity vs. distance to incorporate both parameters. **Plot A** shows the isothermal analytic model **(Section 2.1)** based on equation (5). **Plot B** shows the saturated case where the atmosphere was assumed to follow the Clausius-Clapeyron equation and was saturated from the surface to the critical radius for escape **(Section 2.2)**. **Plot C** shows the isothermal model with the greenhouse effect of water vapor considered **(Section 2.3)**. The results shown in all three figures are dependent on the chosen albedo. If the albedo were to be increased from the chosen value of 0.2, the effect would be a linear decrease in absorbed flux. This would shift the runaway limit and the contours of ocean lifetime closer to the host star. For a Ganymede sized moon with a Bond albedo of 0.2 (shown here), 0.4, and 0.8 the runaway limit occurs at 1.05, 0.91, and 0.52 AU respectively.

## 3. RESULTS

For each body in the range of masses and distances considered, we calculated the time averaged mass loss rate, $\bar{M}$, using a time step of $10^4$ years. With the water content of each world assumed to be 40% of the bulk mass, the water lifetime, $\tau_{Water}$, was then calculated via equation (6). The results of these calculations are shown in Figure 2.

Figure 2 shows contours of $\tau_{Water}$ as a function of stellar distance and escape velocity, which is defined as

$$v_{esc} = \left( \frac{2GM}{r_s} \right)^{1/2} \quad (11)$$

The runaway greenhouse star-exomoon distance is shown with red contours on each plot in Figure 2. From Figure 2A we can see that, in the analytic model, water on a Ganymede-like exomoon (with an escape velocity of ~2.74 km s$^{-1}$) would persist indefinitely at a distance beyond the runaway limit. However, the ice on a Europa sized moon would only survive for timescales greater than 1 Gyr beyond

~1.5 AU. Given that the isothermal and saturated cases represent the upper and lower bounds on escape rate, the true solution is likely somewhere between the two plots (Figures 2A and 2B).

In Figure 2C, the impact of the water vapor greenhouse effect was considered. Under the radiative model, the greenhouse effect of a pure water vapor atmosphere contributes a few degrees K of warming. However, if the body receives sufficient stellar warming a runaway occurs. With a pure water vapor atmosphere the surface never rises above the freezing point of water without entering a runaway greenhouse. But if clouds were to increase the albedo, a world with a liquid water surface may exist with a marginally stable surface temperature up to 275 K (Goldblatt et al. 2013). However, such a world may be transient and easily swing to either a snowball via the ice-albedo feedback, or a runaway greenhouse state (Goldblatt, et al. 2013).

## 4. DISCUSSION

The closely packed lifetime lines in Figure 2 result from a strong dependence on escape velocity and therefore on mass. From equation (5) and (6), with all the constants stripped away, we see that there is an exponential relationship between ocean lifetime and mass, if the mean density of the moon is held constant, given by

$$\tau_{Water} \propto \frac{1}{M} \exp\left[M^{2/3}\right] \qquad (12)$$

For constant density, $\bar{\rho}$, the escape velocity from equation (11) is $v_{esc} = \left(\frac{8}{3}\pi G r_s^2 \bar{\rho}\right)^{1/2} \propto r_s$, so implicit in equation (12), $M \propto v_{esc}^3$ so $\tau_{Water} \propto v_{esc}^{-3} \exp\left(v_{esc}^2\right)$. This strong exponential dependence on $v_{esc}^2$ can be seen in Figure 2 in both the isothermal and saturated cases. There is a threshold mass region, below which surface water is transient, while moons with masses above this region will last for billions of years. Ganymede sized moons will persist indefinitely beyond the star-exomoon distance of the runaway limit.

If a gas giant planet possessed rapidly evaporating icy moons future observations may be able to detect them. The escaped H from water would form a torus in the orbit of the moon that may produce detectable scattering in the Lyman-α. However, for young, migrating planets this H torus may be indistinguishable from captured nebular H before it dissipates. This degeneracy could be addressed by observing aging gas giant planets that are just entering the habitable zone as the host star brightens over time. As a Jupiter-like planet enters the habitable zone around an aging star, hydrogen is unlikely to escape from the planet. Indeed, if we assume a Jupiter-like planet at 0.9 AU around a Sun-like star has an exobase temperature of 1500 K then, following Sánchez-Lavega (2011), p. 88, the thermal loss of hydrogen via Jeans' escape from such a planet will be ~$10^{-37}$ kg s$^{-1}$. This is ~40 orders of magnitude less than the loss rate from icy moons at the same orbital distance so any observed H torus may be an indication of evaporating moons. Icy moons around such a planet are of particular interest because they may provide habitable surface conditions for hundreds of millions to billions of years, depending on the stellar type (Ramirez & Kaltenegger 2016). As the host star brightens the smallest icy moons in the habitable zone would rapidly evaporate, producing the H torus, while more massive moons could retain their surface water for billions of years.

A similar torus-producing process occurs for Io, where a plasma torus around Jupiter contains sulfur and oxygen lost by the moon that are trapped by Jupiter's magnetic field lines (Yoshioka et al. 2011). Also, O atoms may linger around the icy exomoons, analogous to the $O_2$-rich collisional atmosphere of Callisto (Cunningham et al. 2015) and could possibly escape the moon to form a torus similar to the escaped H. A second, heavier component in the exomoon's atmosphere, such as oxygen, would generally act to lower the rate of escape and water loss. However, a more sophisticated model than presented here is required to study escape from a multicomponent atmosphere.

## 5. CONCLUSION

Planetary migration is likely a common phenomenon throughout planetary systems (Tsiganis, et al. 2005). In addition, all the large planets in the solar system have a retinue of icy moons and gas giant exoplanets may have similar icy moons. Inward migration by a gas giant would subject its icy moons to increased stellar heating. Like a comet entering the inner solar system, the moons could evaporate and create atmospheres.

The longevity of such an atmosphere depends strongly on the distance from the host star, and the mass and radius of the exomoon. The smaller the star-exomoon distance, the warmer the icy exomoon will become. As an icy exomoon approaches a distance of ~1.1 AU around a Sun-like star it will enter a runaway greenhouse state when the surface melts. However, this cutoff is dependent on the albedo of the moon, which was set to 0.2 in this paper. Increasing the albedo will allow stable surface conditions at closer orbital distances before the runaway state is achieved. The high temperatures from a runaway state will drive rapid hydrodynamic escape and erode the water from the exomoon on very short timescales.

If the exomoon sits beyond this runaway limit the surface water may persist much longer. Beyond the star-exomoon distance of the runaway limit, there is an exponential relationship between mass and water longevity. For an icy moon of Ganymede's size around a Sun-like star, surface waters will likely persist indefinitely. Large moons of this size will maintain their atmospheres for long periods in the habitable zone and could potentially maintain a liquid surface for timescales greater than 1 Gyr. Thus, such moons could be habitable. However, an icy moon of Europa's size would evaporate rapidly at ~1.1 AU around a Sun-like star, and only beyond ~1.5 AU would surface water (as ice) on a Europa sized moon last for more than 1 Gyr.

**ACKNOWLEDGEMENTS**

ORL, DCC, and KJZ were supported by NASA Planetary Atmospheres grant NNX14AJ45G awarded to DCC. We would like the thank Tyler D. Robinson, for insightful comments and suggestions on this study.

# APPENDIX A: DERIVATION OF ISOTHERMAL AND SATURATED HYDRODYNAMIC ESCAPE MODELS

## A.1. Isothermal Model

The three key equations for hydrodynamic escape – continuity, momentum, and energy – can be written generally (e.g., multiple species, etc. (Koskinen et al. 2013)) but we will use a simplified spherically symmetric model with constant mean molecular mass (see Ch. 5 in (Catling & Kasting 2017) for a more complete discussion of the topic). We assume the atmospheric density and atmospheric flow velocity only change in the radial direction. As such, the derivatives for mass continuity and momentum conservation are complete. Under these assumptions, the time-dependent and steady-state continuity and mass conservation equations are as follows. Continuity is given by:

$$\frac{\partial \rho}{\partial t} = -\frac{1}{r^2}\frac{d}{dr}\left(r^2 \rho u\right) \quad , \text{steady state:} \quad \frac{d}{dr}\left(r^2 \rho u\right) = 0 \qquad (A1)$$

where $\rho$ is the mass density, $r$ is the radial distance from the planet's center, and $u$ is the atmospheric flow velocity. Momentum conservation is given by:

$$\frac{\partial (\rho u)}{\partial t} = -\rho u \frac{du}{dr} - \frac{dp}{dr} + \rho g \quad , \text{steady state:} \quad u\frac{du}{dr} + \frac{1}{\rho}\frac{dp}{dr} = g \qquad (A2)$$

where $p$ is pressure, and $g$ is gravity.

If we assume an isothermal atmosphere, we can relate pressure and density with the isothermal sound speed

$$u_0^2 = \frac{kT}{m} \qquad (A3)$$

where $k$ is the Bolztmann constant, $T$ is the isothermal temperature, and $m$ is the mean molecular mass of the atmosphere. From the ideal gas law

$$p = \rho u_0^2 \qquad (A4)$$

Integrating equation (A1) in the steady state, we get the mass escape rate per steradian of $r^2 \rho u$, which, when combined with equations (A1) and (A2), gives an expression for the isothermal planetary wind from a body with mass $M$:

$$\left(u^2 - u_0^2\right)\frac{1}{u}\frac{du}{dr} = \frac{2u_0^2}{r} + g \quad, \text{ or } \quad \left(u^2 - u_0^2\right)\frac{1}{u}\frac{du}{dr} = \frac{2u_0^2}{r} - \frac{GM}{r^2} \tag{A5}$$

Equation (A5) is analogous to Parker's solar wind equation. For a strongly bound atmosphere at some critical distance from the planet's surface, the right hand side of equation (A5) reaches zero, indicating that either the flow reaches the speed of sound or $(du/dr)_c = 0$. The subsonic solution, $(du/dr)_c = 0$, requires a finite background pressure that inhibits escape so we will focus on the transonic solution where $u^2 = u_0^2$. The transonic solution has $du/dr > 0$ at all times and is consistent with a strongly bound atmosphere at the surface and zero pressure at infinity.

The critical distance $r_c$ occurs in equation (A5) when $u^2 = u_0^2$ which gives us:

$$0 = \frac{2u_0^2}{r_c} - \frac{GM}{r_c^2} \tag{A6}$$

Solving for $r_c$ in equation (A6) we find

$$r_c = \frac{GM}{2u_0^2} \tag{A7}$$

If we integrate equation (A5) from the surface radius $r_s$ to $r_c$ and ignore the $u^2$ term near the surface, where it is negligible for bodies of interest in this study, we get the equation:

$$\rho_s u_s = \rho_s u_0 \left(\frac{r_c}{r_s}\right)^2 \exp\left[-\frac{1}{2} + \frac{GM}{u_0^2}\left(\frac{1}{r_c} - \frac{1}{r_s}\right)\right] \tag{A8}$$

As the radial distance from the moon increases the mass flux, $\rho u$, (in kg m$^{-2}$ s$^{-1}$), decreases. The steady state continuity given by equation (A1), when integrated gives $4\pi r^2 \rho u = C$ where the constant of integration $C$ is just the total rate of mass loss (in kg s$^{-1}$) through a spherical surface. As $r$ goes to infinity $\rho u$ goes to 0 since $\rho u \propto 1/r^2$. Therefore, the outflowing wind loses kinetic energy as $r \to \infty$. Thus, the energy flux required to drive the escaping mass flux is given by the energy required to remove the mass flux from the gravity well of the moon, $\rho_s u_s GM / r_s$.

A first-order global energy balance between insolation and cooling via mass loss is then given by:

$$\underbrace{\frac{1}{4}(1-A)F_s}_{\text{absorbed stellar flux}} = \underbrace{\left(\frac{GM}{r_s} + L_v\right)\rho_s u_s}_{\text{mass loss flux}} + \underbrace{\sigma T^4}_{\text{radiative cooling}} \tag{A9}$$

where $A$ is the Bond albedo, $F_s$ is the incident stellar flux, and $\sigma$ is the Stefan-Boltzmann constant. The escape flux is given by $\rho_s u_s$ and is multiplied by the energy required for that flux to escape the planet. The energy includes a gravitational potential energy term, and the latent heat of vaporization $L_v$ (for this model $L_v = 2.5 \times 10^6$ J kg$^{-1}$). In equation (A9) we assume the atmosphere is transparent to both shortwave and infrared radiation.

Equations (A8), and (A9) can be solved simultaneously for the two unknowns $u_s$ and $T$. Once solved, we can calculate the total escaping mass rate by:

$$\dot{M} = 4\pi \rho_s u_s r_s^2 \tag{A10}$$

with $\rho_s$ being calculated from $\rho_s = P_s / u_0^2$ with surface pressure $P_s$. We calculate surface pressure with equation (1) for a C-C atmosphere given the surface temperature of water, where reference parameters are at the triple point: $P_0 = 611.73$ Pa, $T_0 = 273.16$ K. For our model, we only consider water worlds with pure H$_2$O atmospheres so estimating the surface density from the saturation vapor pressure is valid (Adams, Seager, & Elkins-Tanton 2008). We refer to this approach, where equations (A8) and (A9) are solved numerically, as the **Numerical Model**.

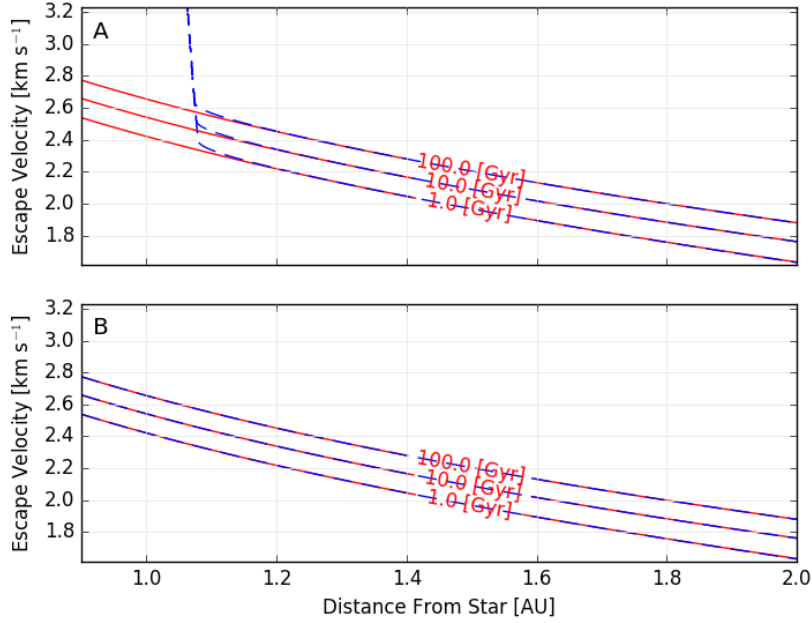

**Figure A1.** Contours of surface water lifetime comparing the analytic model given by equation (5), shown in red, and the numerical approach where $T$ and $u_s$ are solved for simultaneously, shown in dashed blue contours. **Plot A** shows the analytic model, which does not consider the greenhouse effect, plotted with the numerical model taking into account the greenhouse effect of water vapor as derived in **Appendix B**. Both models produce identical results until the runaway limit is approached and the numerical model asymptotes along the limit. **Plot B** shows the analytic and numerical models as well; however, the greenhouse effect is neglected in the numerical model for this plot. In this case, both methods produce identical results, as expected, for slowly evaporating bodies with surface water lasting more than 1 Gyr.

For the slowly evaporating moons of interest in this study, those with surface water lasting more than 1 Gyr, the escape is so slow it does not appreciably cool the moon. Thus an analytic model can be derived by neglecting the mass loss flux cooling term in equation (A9). With the simplified equation (A9) our isothermal temperature is simply calculated from incoming stellar flux. And, from our assumption that the exomoons have an average bulk density of 2 g cm$^{-3}$, we can calculate the surface radius $r_s = (3M / (4\pi \rho_{exomoon}))^{1/3}$. Substituting these two equations into equation (A8) we find that

$$u_s = u_0 \frac{r_c^2}{r_s^2} \exp\left[ \frac{3}{2} - \frac{G}{u_0^2} \left( \frac{3}{4\pi \rho_{exomoon}} \right)^{\frac{-1}{3}} M^{2/3} \right] \quad (A11)$$

By plugging equation (A11) into equation (A10), we get the analytic expression for the mass loss due to hydrodynamic escape given in equation (5).

Calculating $u_s$ in this manner assumes the temperature in our energy balance equation is a constant and set solely by the incoming stellar flux and the emitted thermal flux from the surface. For the low temperature bodies ($< 273$ K) we are interested in for this study, equation (A11) gives identical results as the previously defined numerical model until the runaway limit is approached. See Figure A1 for a comparison.

### A.2. Saturated Model

We also modeled hydrodynamic escape from a non-isothermal atmosphere, the saturated case. To model escape in the saturated case we start with equations (A1), (A2), (A3), and (A4). Instead of using equation (1) to relate temperature and pressure, we will approximate the Clausius-Clapeyron relation with an expression similar to Tentens' formula, given by

$$p = p_w \exp(-T_w / T) \quad (A12)$$

for reference temperature $T_w$ and pressure $p_w$. A reasonable approximation for $250 < T < 400$ K over water takes $T_w = 5200$ K and $p_w = 1.13 \times 10^6$ bar. A very good approximation for $150 < T < 273$ K over ice takes $T_w = 6140$ K and $p_w = 3.53 \times 10^7$ bar. From Wexler (1977), whose expression we've approximated, the simple exponential fit is likely good to within a few percent for the temperatures in our model. This simplified expression is desirable because we want to work with an analytic expression for $dT/dr$.

We can eliminate $p$ from equation (A2) using equations (A3) and (A4), giving us

$$u\frac{du}{dr} + \frac{u_0^2}{\rho}\frac{d\rho}{dr} + \frac{u_0^2}{T}\frac{dT}{dr} = -\frac{GM}{r^2} \quad (A13)$$

We use equation (A12) to express $dT/dr$ in terms of $d\rho/dr$

$$\frac{u_0^2}{\rho}\frac{d\rho}{dr} = \left(\frac{T_w - T}{T}\right)\frac{u_0^2}{T}\frac{dT}{dr} \quad (A14)$$

and equation (A1) eliminates $d\rho/dr$ in terms of $du/dr$ giving us our saturated wind equation

$$\left(u - \frac{u_0^2}{u}\left(\frac{T_w}{T_w - T}\right)\right)\frac{du}{dr} = \frac{2u_0^2}{r}\left(\frac{T_w}{T_w - T}\right) - \frac{GM}{r^2} \quad (A15)$$

or equivalently as an expression for $du/dr$

$$\frac{1}{u}\frac{du}{dr} = \frac{N}{D} = \frac{(2u_0^2/r)(T_w/(T_w - T)) - GM/r^2}{u^2 - u_0^2(T_w/(T_w - T))} \quad (A16)$$

where the numerator $N(r,T)$ is

$$N = \frac{2u_0^2}{r}\left(\frac{T_w}{T_w - T}\right) - \frac{GM}{r^2} \quad (A17)$$

and the denominator $D(r,T,u)$ is

$$D = u^2 - u_0^2\left(\frac{T_w}{T_w - T}\right) \quad (A18)$$

Equation (A16) is the form we will use to numerically integrate $u(r)$. Equation (A15) can be written equivalently as

$$D\frac{1}{u}\frac{du}{dr} = N \quad (A19)$$

Recall from the isothermal case that, for hydrodynamic escape from a strongly bound atmosphere, $du/dr > 0$. Near the surface of the moon the numerator, $N(r,T)$, will be negative as the gravity term will dominate given that our atmosphere is strongly bound. At some distance $r_c$ the $2u_0^2/r$ term will equal the force of gravity, so $N(r,T) = 0$ at $r_c$. Since $N(r,T) = 0$ and $du/dr > 0$, from equation (A19), $D(r,T,u) = 0$ at $r_c$ as well. At the critical point, $N_c = 0$ provides a simple relation between $T_c$ and $r_c$

$$r_c = \frac{GMm(T_w - T_c)}{2kT_c T_w} \tag{A20}$$

Similarly, $D_c = 0$ relates $u_c$ and $T_c$ by

$$u_c^2 = \left(u_0^2\right)_c \left(\frac{T_w}{T_w - T_c}\right) = \frac{GM}{2r_c^3} \tag{A21}$$

The transonic solution is obtained by numerically integrating equation (A16) from the critical point to the surface. The first step is to solve for $(du/dr)_c$ at the critical point. This is obtained from equation (A16) by using L'Hopital's rule.

$$\frac{1}{u_c}\left(\frac{du}{dr}\right)_c = \frac{N_c}{D_c} = \frac{0}{0} = \frac{(dN/dr)_c}{(dD/dr)_c} \tag{A22}$$

The numerator becomes

$$\left(\frac{dN}{dr}\right)_c = \frac{GM}{r_c^3} - \frac{4\left(u_0^2\right)_c}{r_c^2}\frac{T_w^2 T_c}{(T_w - T_c)^3} - \frac{2\left(u_0^2\right)_c}{r_c}\frac{T_w^2 T_c}{(T_w - T_c)^3}\frac{1}{u_c}\left(\frac{du}{dr}\right)_c \tag{A23}$$

and the denominator becomes

$$\left(\frac{dD}{dr}\right)_c = \left(2 + \frac{T_w T_c}{(T_w - T_c)^2}\right) u_c \left(\frac{du}{dr}\right) + \frac{2u_c^2}{r_c^2} \frac{T_w T_c}{(T_w - T_c)^2} \tag{A24}$$

If we let $x \equiv u_c^{-1}(du/dr)_c$, simplify equation (A23) to replace $GM/r_c$ with equation (A21), and divide out the common factor $u_c^2$, then equation (A22) can be written as the quadratic equation

$$\left(2 + \frac{T_w T_c}{(T_w - T_c)^2}\right) x^2 + \frac{4}{r_c} \frac{T_w T_c}{(T_w - T_c)^2} x + \frac{4}{r_c^2} \frac{T_w T_c}{(T_w - T_c)^2} - \frac{2}{r_c^2} = 0 \tag{A25}$$

The positive root of this equation corresponds to the accelerating flow at the critical point.

To find the mass flux loss, the first step is to guess an initial temperature at the critical point, $T_c$. Given $T_c$, we know $u_c^2$, $p_c$ is given from equation (A12), and $\rho_c$ is given by the ideal gas law. From equations (A20) and (A21), we get $r_c$ and $u_c$ respectively, which allows us to solve equation (A25) for the critical slope $(du/dr)_c$. Density can then be found at the new point from continuity, $\rho u r^2 = \rho_c u_c r_c^2$. Given $\rho$, we can solve for $T$ and $p$ from equation (A12) with the help of the ideal gas equation. This integration proceeds to the surface. The guess for $T_c$ is adjusted numerically until the desired surface temperature (in balance with incoming stellar flux and mass loss given by equation (A9)) is achieved. Once the correct values are found, equation (A10) will give the mass loss rate.

For both isothermal and non-isothermal models, the surface temperature is assumed to be set by the incident solar flux averaged over time and hemisphere, which is given by equation (2) for a rapidly rotating body. The isothermal case represents the warmest possible atmosphere neglecting greenhouse effects under the case of hydrodynamic escape. The non-isothermal case represents a minimum possible temperature for a water vapor atmosphere at $r_c$ since it is saturated at all points based on the surface temperature set from the solar flux. These two models represent the extremes of atmospheric temperature profiles for a water vapor atmosphere, with the real solution likely somewhere between them.

**APPENDIX B: DERIVATION OF SURFACE TEMPERATURE ACCOUNTING FOR GREENHOUSE EFFECT AND HYDRODYNAMIC ESCAPE**

We would like to calculate the total surface warming due to the greenhouse effect of a water vapor atmosphere considering the energy absorbed to drive atmospheric expansion and escape throughout the atmosphere. We start with the greenhouse effect of a hydrostatic atmosphere, then adapt the equation for a hydrodynamic atmosphere. We assume the atmosphere is transparent to shortwave radiation. From Catling and Kasting (2017), p. 55, for a moon with a gray, radiative, hydrostatic atmosphere the energy balance at the surface is given by

$$\sigma T_s^4 = F_{net}\left(1 + \tau/2\right) \tag{B1}$$

where $\tau$ is the total thermal infrared optical depth of the atmosphere at the surface, $\sigma$ is the Stefan-Boltzmann constant, and $T_s$ is the surface temperature. The time-averaged, hemispherically-averaged flux incident on the moon is given by $F_{net} = (1-A)F_s/4$ for Bond albedo $A$, and incident stellar flux $F_s$.

In our model, we are concerned with moons in the hydrodynamic regime where water vapor is lifted from the surface of the moon and accelerates upward until it escapes to space. The total energy required to remove a mass flux of water vapor from the moon's surface is given by

$$\frac{GM}{r_s}\rho_s u_s \tag{B2}$$

for surface radius $r_s$. In equation (B2) $M$ is the mass of the moon, $G$ is the gravitational constant, $u_s$ is the radial outflow velocity of the atmosphere at the surface, and $\rho_s$ is the atmospheric density at the surface, such that $r_s u_s$ is the mass flux [kg m$^{-2}$ s$^{-1}$].

In the hydrodynamic atmospheres of interest in this study, the energy flux needed to remove the atmosphere, given by equation (B2), must come from the stellar radiation and the thermal IR flux. That is, it must come from the $F_{net}(1+\tau/2)$ energy input term in equation (B1). As such, the energy balance at the surface will then be given by

$$\sigma T_s^4 = F_{net}(1+\tau/2) - \frac{GM}{r_s}\rho_s u_s \quad (B3)$$

in the hydrodynamic regime. We also account for the energy required to vaporize the water mass flux at the surface, given by $L_v \rho_s u_s$ for latent heat of vaporization $L_v$. Subtracting $L_v \rho_s u_s$ from the right-hand side of equation (B3) and reorganizing the terms we find the following energy balance of input and output:

$$\frac{1}{4}(1-A)F_s\left(1+\frac{\tau}{2}\right) = \left(\frac{GM}{r_s}+L_v\right)\rho_s u_s + \sigma T_s^4 \quad (B4)$$

Equation (B4) is the global energy balance at the surface for an icy moon with the greenhouse effect considered under the hydrodynamic regime. It can be compared with equation (2) in the main text where we assumed an atmosphere that was optically thin in the thermal infrared.